\documentclass[lettersize,journal]{IEEEtran}
\usepackage{cite}
\usepackage{amsmath,amssymb,amsfonts}
\usepackage{algorithmic}
\usepackage{graphicx}
\usepackage{textcomp}
\usepackage{xcolor}
\usepackage{hyperref}
\usepackage{enumitem}
\usepackage{subfigure}
\usepackage{booktabs}
\usepackage{multirow}

\def\BibTeX{{\rm B\kern-.05em{\sc i\kern-.025em b}\kern-.08em
    T\kern-.1667em\lower.7ex\hbox{E}\kern-.125emX}}
\begin{document}

\DeclareRobustCommand{\loongrightarrow}{%
  \DOTSB\relbar\joinrel\relbar\joinrel\rightarrow
}
\DeclareRobustCommand{\loongmapsto}{\DOTSB\mapstochar\loongrightarrow}

\DeclareRobustCommand{\looongrightarrow}{%
  \DOTSB\relbar\joinrel\relbar\joinrel\relbar\joinrel\rightarrow
}
\DeclareRobustCommand{\looongmapsto}{\DOTSB\mapstochar\looongrightarrow}

\title{{\fontsize{20pt}{17pt}\selectfont CORF-GS: Real-Time Wireless Radiance Field Reconstruction via Coupled Optical-RF Gaussian Splatting}}
\author{Jinya~Zhang,~Jiajia~Guo,~Chao-Kai~Wen, \IEEEmembership{\normalsize{Fellow,~IEEE}}, and Shi~Jin, \IEEEmembership{\normalsize{Fellow,~IEEE}}}

\maketitle

\begin{abstract}
Recent advances in 3D Gaussian Splatting (3DGS)-based wireless radiance field (WRF) reconstruction provide an efficient solution for wireless channel modeling. However, existing WRF reconstruction methods rely on pre-collected observations and offline optimization, and thus struggle to provide real-time channel knowledge. To bridge this gap, we propose CORF-GS, a real-time WRF reconstruction framework that processes sequential optical and radio frequency (RF) keyframes. Specifically, CORF-GS constructs a unified Gaussian representation for optical and RF with shared geometry and modality-specific appearance, allowing high-resolution optical images to provide structural priors for WRF reconstruction. When a new keyframe arrives, CORF-GS first employs optical-guided Gaussian sampling to densify the WRF in under-represented regions. Since light and radio waves may respond differently to the same object surfaces due to wavelength mismatch, relying solely on optical guidance may neglect RF-informative areas. Therefore, CORF-GS performs coupled optical-RF optimization to jointly refine the shared Gaussians. Compared with the existing two-stage training pipelines, this prevents WRF from passively adapting to a frozen optical geometry and encourages the shared Gaussians to adapt to both optical structures and RF power distributions. Simulations show that CORF-GS achieves state-of-the-art RF spectrum synthesis quality and reduces the reconstruction time by $6.4\times$ compared with existing WRF methods.
\end{abstract}

\section{Introduction}
Future wireless networks are expected to support emerging applications such as virtual reality, wireless digital twins, and seamless sensing~\cite{mihai2022digital,liu2022integrated}. These applications require the deep integration of communication, sensing, computing, and intelligence, imposing stringent demands on data rate, latency, and robustness~\cite{itur2023imt2030}. To meet these demands, wireless systems need an accurate understanding of the interaction between radio waves and surroundings to acquire channel knowledge. Therefore, wireless optimization requires dense, efficient, and accurate wireless channel modeling.

Existing wireless channel modeling techniques still face limitations in meeting these requirements. Probabilistic models, such as the 3GPP channel model, describe channel characteristics using statistical distributions extracted from measurements~\cite{sarkar2003survey}. They are efficient but lack spatial details, making it difficult to support fine-grained channel acquisition. Deterministic methods, such as ray tracing~\cite{he2018design,orekondy2023winert}, simulate electromagnetic wave propagation based on a digital environment model. Although they can provide accurate predictions, their performance depends heavily on precise geometry and material properties, and their computational cost is often high. Neural network (NN)-based methods~\cite{levie2021radiounet} directly learn the mapping from environmental information to channel properties from data. However, they commonly formulate channel modeling as a black-box regression problem, which limits their robustness under sparse measurements.

Recent advances in neural rendering, especially neural radiance fields (NeRF)~\cite{mildenhall2020nerf} and 3D Gaussian Splatting (3DGS)~\cite{kerbl20233d}, have provided a new perspective for wireless channel modeling. Originally developed for optical scene reconstruction, these methods also provide physics-informed representations for wireless radiance fields (WRF)~\cite{sun2026radio}. This connection is motivated by the common electromagnetic nature of light and radio waves, which both interact with the environment through reflection, attenuation, and blockage. Along this direction, NeRF-based methods parameterized WRF with NNs, enabling dense radio frequency (RF) spectrum synthesis from limited observations~\cite{zhao2023nerf2,lu2024newrf,wang2026sign}. However, their implicit representation requires costly optimization. In contrast, 3DGS uses explicit Gaussian primitives and efficient differentiable splatting, making it a more promising method for fast and accurate WRF reconstruction~\cite{wen2025neural,xue2026point}.

Nevertheless, two key challenges still hinder the practical deployment of 3DGS-based WRF reconstruction. First, existing methods predominantly construct the WRF in an offline and static manner. Consequently, the WRF may fail to reflect the real-time propagation condition, leading to unreliable channel acquisition. Second, several frameworks adopt a two-stage pipeline that first optimizes Gaussian geometry from optical images and then freezes it for RF-supervised appearance optimization. Although this strategy leverages rich geometric priors, it assumes that the optical Gaussian structure can well support WRF reconstruction. This assumption is not always valid, since the propagation of light and radio waves is dominated by different physical attributes due to their substantial wavelength discrepancy. Therefore, Gaussian geometry may overemphasize optical textures while underrepresenting RF-related macro-structures, degrading the modeling of energy peaks and multi-path distributions.

To address these challenges, we propose CORF-GS, a 3DGS-based real-time WRF reconstruction framework with sequential optical-RF observations. The contributions are summarized as follows:
\begin{itemize}[leftmargin=*]
    \item We introduce a coupled optical-RF Gaussian representation, where optical and RF share the same Gaussian geometry while maintaining modality-specific appearance. This design provides common 3D geometric support for both optical radiance field (ORF) and WRF reconstruction, enabling WRF to be associated with scene structures.
    \item We propose a real-time WRF reconstruction framework from sequential optical-RF observations under known-pose settings. CORF-GS samples new primitives from optical structures upon the arrival of each keyframe, bypassing the time-consuming gradient-based densification. To the best of our knowledge, CORF-GS is the first attempt to reconstruct a 3DGS-based WRF in real time.
    \item We develop an RF-aware coupled optimization strategy to mitigate the discrepancy between optical and RF propagation. Instead of forcing the WRF to passively fit frozen optical geometry, CORF-GS jointly refines the shared Gaussians with both optical and RF supervision. This design balances optical structural constraints and RF spectrum consistency for more accurate WRF reconstruction.
\end{itemize}

\section{Related Work}
\subsection{3DGS-Based WRF Reconstruction}
The key idea of 3DGS-based WRF reconstruction is to reinterpret 3D Gaussian primitives as RF-relevant elements to render RF spatial (angular) spectra at receivers (RXs), which characterize multi-path power distributions. Although both light and radio waves are governed by Maxwell's equations~\cite{yun2015ray}, their wavelength and propagation differences require RF-specific adaptations in primitive semantics, projection models, and optimization strategies.

Existing 3DGS-based WRF methods generally fall into two categories based on how they model the interaction between radio waves and the propagation environment. 
The first category interprets Gaussians as radio-propagating elements and purely optimizes RF attributes from radio measurements. For instance, WRF-3DGS~\cite{wen2025wrf} represented Gaussian primitives as virtual transmitters (TXs) to model wave-environment interactions, utilizing a Mercator projection model to map hemispherical observations into angular-domain spectra. WRF-GS+~\cite{wen2025neural} further extended this framework with deformable Gaussians to decouple static and dynamic field components. Different from NeRF-style formulations that rely on NNs, GSRF~\cite{yang2026gsrf} directly optimized RF-specific Gaussian attributes to formulate spatial spectra. GSpaRC~\cite{nukapotula2025gsparc} further improved reconstruction efficiency by developing a fast WRF reconstruction framework. Other works leverage optical observations to provide scene priors for WRF reconstruction. RF-3DGS~\cite{zhang2026rf} first reconstructed the scene structure from optical images and then froze the Gaussian geometry to learn power distribution from spatial spectra. Similarly, URF-GS~\cite{wen2026bridging} adopted a two-stage optimization pipeline but focused on recovering underlying material electromagnetic properties rather than RF appearance, thus improving spectra synthesis accuracy.

However, existing methods mainly rely on offline reconstruction from pre-collected observations. Optical-assisted methods force the WRF to fit a frozen, optically optimized geometry, which may overlook RF-informative regions due to wavelength-dependent response discrepancies.

\subsection{Real-Time 3D Gaussian Reconstruction}
Standard 3DGS~\cite{kerbl20233d} and its variants~\cite{feng2025flashgs,zhao2025scaling,mallick2024taming,ye2025gsplat,fan2024instantsplat} are mainly designed for offline reconstruction, limiting their practicality for real-time tasks. Therefore, recent studies have explored real-time Gaussian reconstruction, where a key challenge is how to efficiently expand the Gaussian representation as new observations arrive.

GS-SLAM~\cite{Yan_2024_CVPR} introduced an adaptive Gaussian expansion strategy based on opacity and depth residuals, selecting unreliable pixels and back-projecting them to a 3D point. 
Photo-SLAM~\cite{huang2024photo} maintained a hyper primitives map and improved online mapping quality through geometry-based densification and pyramid-based learning. 
SplaTAM~\cite{keetha2024splatam} expanded the map to insert new Gaussians in regions with insufficient silhouette coverage or inconsistent depth observations. 
RTG-SLAM~\cite{peng2024rtg} added Gaussians for newly observed pixels with large color or depth errors, while optimizing only unstable Gaussians.
On-The-Fly NVS~\cite{meuleman2025fly} further proposed a probability-based direct Gaussian sampling strategy, placing primitives to cover previously unseen regions or to add additional details. 

While existing real-time methods adaptively insert Gaussians into under-represented regions, they are strictly optimized for optical scenes. When extending to optical-RF reconstruction, the core challenge escalates to accommodate both optical appearances and multi-path distribution, which differ in physical characteristics.

\section{Preliminaries}

\begin{figure*}[t]
    \centering
    \includegraphics[width=0.98\textwidth]{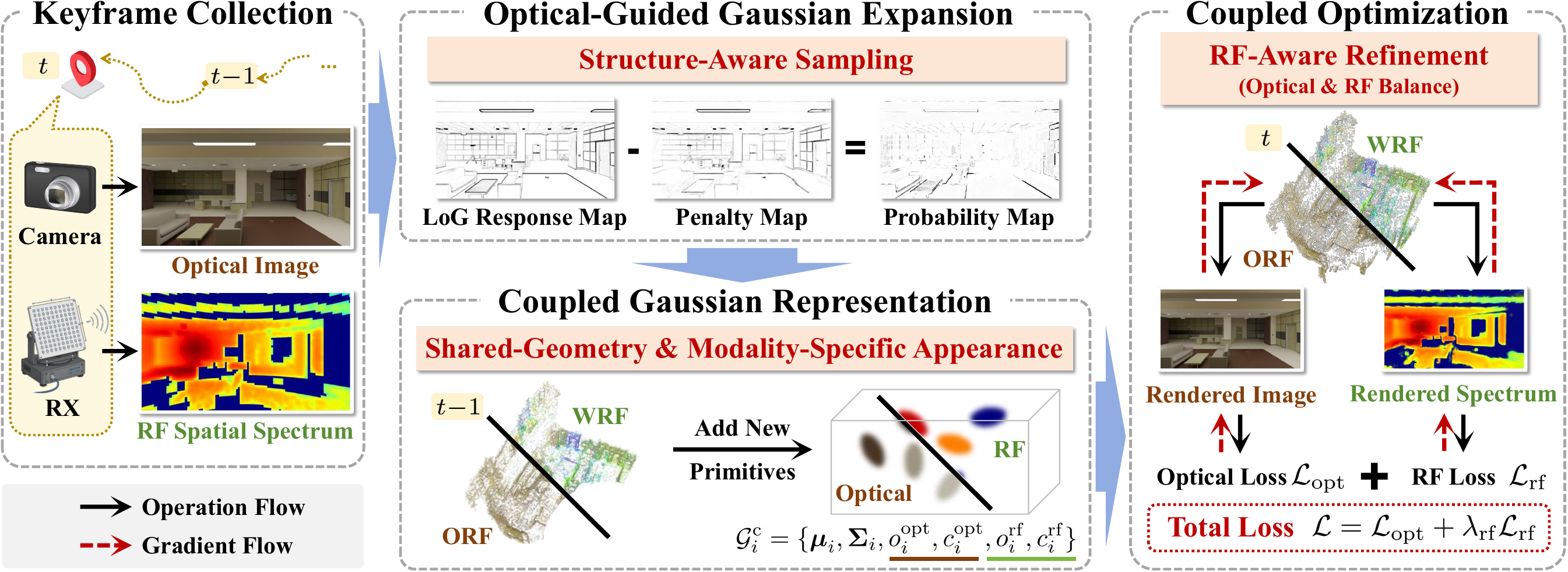}
    \caption{The workflow of CORF-GS. CORF-GS constructs a coupled Gaussian representation for optical and RF with shared geometry and modality-specific appearance. 
    When a new keyframe arrives, CORF-GS first employs optical-guided Gaussian expansion in under-represented regions. Since relying solely on optical guidance may neglect RF-informative areas, CORF-GS then performs coupled optical-RF optimization to jointly refine the shared Gaussians.}
    \label{fig:workflow}
\end{figure*}

\subsection{Wireless Channel Modeling} 
In wireless communications, RF signals propagate from a TX to an RX through interacting with the surrounding environment. Formally, the wireless channel is characterized as a superposition of multi-path components:
\begin{equation}
    \mathbf{H}(\tau) = \sum_{l=1}^{L} \beta_{l} \mathbf{a}_\mathrm{r}(\Omega^\mathrm{r}_{l}) \mathbf{a}_\mathrm{t}^{H}(\Omega^\mathrm{t}_{l}) \delta(\tau-\tau_{l}),
\end{equation}
where $L$ is the number of propagation paths; $\beta_{l} \in \mathbb{C}$ and $\tau_{l}$ denote the path gain and delay of the $l$-th path, respectively. The vectors $\mathbf{a}_\mathrm{r}(\Omega^\mathrm{r}_{l})$ and $\mathbf{a}_\mathrm{t}(\Omega^\mathrm{t}_{l})$ represent the RX and TX antenna array responses, parameterized by the angle-of-arrival (AoA) $\Omega^\mathrm{r}_{l}$ and angle-of-departure (AoD) $\Omega^\mathrm{t}_{l}$.

Conventional channel modeling relies on exhaustive grid-based measurements across the 3D space. In contrast, the efficient WRF $\mathcal{F}_\mathrm{rf}$ represents propagation contributors and models power distribution with RF-specific Gaussians.
When querying $\mathcal{F}_\mathrm{rf}$ at a given RX pose $\mathbf{P}_\mathrm{r}$, a spatial spectrum $\hat{\mathbf{S}}$ is rendered by $\mathcal{R}_\mathrm{rf}$ to characterize the angular power distribution of the received signals:
\begin{equation}
    \hat{\mathbf{S}}(\mathbf{P}_\mathrm{r}) = \mathcal{R}_\mathrm{rf}(\mathcal{F}_\mathrm{rf};\mathbf{P}_\mathrm{r}).
\end{equation}
Here, $\mathcal{R}_{\mathrm{rf}}$ follows the same differentiable splatting process as optical 3DGS. Each RF spectrum can be preprocessed into a perspective RF image compatible with the 3DGS camera model~\cite{zhang2026rf}.

\subsection{3D Gaussian Splatting}
3DGS reconstructs a scene using a set of optimizable, anisotropic 3D Gaussian primitives, enabling efficient and high-fidelity novel view synthesis. The ORF is reconstructed with $N$ Gaussians, and each primitive $\mathcal{G}_i$ is parameterized by its spatial center $\boldsymbol{\mu}_i \in \mathbb{R}^{3}$, covariance matrix $\boldsymbol{\Sigma}_i \in \mathbb{R}^{3 \times 3}$, opacity $o_i$, and view-dependent color $c_i$:
\begin{equation}
    \mathcal{G}_i = \{\boldsymbol{\mu}_i, \boldsymbol{\Sigma}_i, o_i, c_i\}, \mathcal{F}_\mathrm{opt} \triangleq \{\mathcal G_i\}_{i=1}^N.
\end{equation}
The spatial distribution of the $i$-th Gaussian at a 3D position $\mathbf{x} \in \mathbb{R}^3$ is defined as:
\begin{equation}
    G_i(\mathbf{x}) = \exp \left( -\frac{1}{2} (\mathbf{x}-\boldsymbol{\mu}_i)^\top \boldsymbol{\Sigma}_i^{-1} (\mathbf{x}-\boldsymbol{\mu}_i) \right).
\end{equation}

For rendering, these 3D Gaussians are projected onto the 2D image plane given a specific camera pose $\mathbf{P}_\mathrm{c}$, and the rendering operator $\mathcal{R}_\mathrm{opt}(\cdot)$ synthesizes the image $\hat{\mathbf{I}}$:
\begin{equation}
    \hat{\mathbf{I}}(\mathbf{P}_\mathrm{c}) = \mathcal{R}_\mathrm{opt}(\mathcal{F}_\mathrm{opt}; \mathbf{P}_\mathrm{c}),
\end{equation}
where the color of a target pixel $\mathbf{u} \in \hat{\mathbf{I}}$ is accumulated via $\alpha$-blending:
\begin{equation}
    C(\mathbf{u}) = \sum_{i=1}^{N} \alpha_i(\mathbf{u}) c_i \prod_{j=1}^{i-1} \left(1-\alpha_j(\mathbf{u})\right).
\end{equation}
Here, $\alpha_i(\mathbf{u}) = o_i g_i(\mathbf{u})$, and $g_i(\mathbf{u})$ denotes the projected 2D Gaussian weight evaluated at pixel $\mathbf{u}$. The Gaussians are optimized by minimizing the reconstruction error between the rendered views and ground-truth images.

\section{Method}
\subsection{Problem Formulation}
The primary objective of CORF-GS is real-time incremental reconstruction of the WRF $\mathcal{F}_{\mathrm{rf}}$ from sequential optical-RF observations, while concurrently maintaining the ORF $\mathcal{F}_{\mathrm{opt}}$ to provide structural guidance. Specifically, we consider a scenario where a co-located camera-RX antenna array platform moves along a continuous trajectory. At each time step $t$ along this trajectory, an optical image $\mathbf{I}_t$ and an RF spatial spectrum $\mathbf{S}_t$ are synchronously captured at the same pose $\mathbf{P}_t = \mathbf{P}_\mathrm{c} = \mathbf{P}_\mathrm{r}$, constituting a keyframe:
\begin{equation}
    \mathcal{K}_t = \{\mathbf{I}_t, \mathbf{S}_t, \mathbf{P}_t\}.
\end{equation}
To simplify the problem formulation, we assume that the keyframe poses are directly obtainable from the device metadata. Upon the arrival of $\mathcal{K}_t$, CORF-GS recursively updates the dual-field representation:
\begin{equation}
    \left( \mathcal{F}_{\mathrm{opt}}^{(t)}, \mathcal{F}_{\mathrm{rf}}^{(t)} \right) = \mathcal{U} \left( \mathcal{F}_{\mathrm{opt}}^{(t-1)}, \mathcal{F}_{\mathrm{rf}}^{(t-1)}; \mathcal{K}_t \right),
\end{equation}
where $\mathcal{U}(\cdot)$ denotes the real-time field update operator. Specifically, $\mathcal{U}(\cdot)$ expands the Gaussian primitive set and refines their attributes to accommodate the incoming keyframe.

After processing a sequence of $T$ keyframes, CORF-GS converges to a locally queryable dual-field representation:
\begin{equation}
    \left( \mathcal{F}_{\mathrm{opt}}^{(T)}, \mathcal{F}_{\mathrm{rf}}^{(T)} \right) = \mathcal{U}_{1:T} \left( \mathcal{K}_1, \mathcal{K}_2, \ldots, \mathcal{K}_T \right).
\end{equation}
Given an arbitrary query pose $\mathbf{P}_\mathrm{q}$ within the reconstructed region, both the optical view $\hat{\mathbf{I}}_\mathrm{q}$ and the RF spatial spectrum $\hat{\mathbf{S}}_\mathrm{q}$ can be synthesized via:
\begin{equation}
    \hat{\mathbf{I}}_\mathrm{q} = \mathcal{R}_\mathrm{opt} \left( \mathcal{F}_{\mathrm{opt}}^{(T)}; \mathbf{P}_\mathrm{q} \right), \quad 
    \hat{\mathbf{S}}_\mathrm{q} = \mathcal{R}_\mathrm{rf} \left( \mathcal{F}_{\mathrm{rf}}^{(T)}; \mathbf{P}_\mathrm{q} \right).
\end{equation}

\subsection{Overview of CORF-GS}
Given that radio waves interact with object surfaces through mechanisms such as reflection and scattering, we assume that radio radiance originates from these surfaces, which are represented by Gaussian primitives. Therefore, in CORF-GS, the ORF and WRF are built upon coupled Gaussian primitives, where the two modalities share the same Gaussian geometry but maintain modality-specific appearance attributes. When a new keyframe arrives, CORF-GS performs three steps:

\begin{itemize}
    \item \textbf{Optical-Guided Gaussian Expansion.} To cover newly observed or under-reconstructed regions, CORF-GS first identifies informative regions from the optical image $\mathbf{I}_t$ and determines pixels for spawning new Gaussian primitives. Compared with gradient-based Gaussian densification, the structure-aware sampling mechanism is more efficient for real-time WRF reconstruction.

    \item \textbf{Parameter Initialization.} Following candidate pixel sampling, CORF-GS back-projects these 2D pixels into 3D space with the assistance of depth estimates and sampling probability priors. Since optical images offer higher-fidelity details than spectra, the initial distribution of the Gaussians is predominantly determined by the optical modality to establish a reliable geometric foundation.

    \item \textbf{Coupled Optical-RF Optimization.} Finally, CORF-GS jointly refines the shared Gaussian geometry and modality-specific appearances under both optical and RF supervision. 
    This process adopts coarse-to-fine optimization, followed by an optional offline global fine-tuning stage after all keyframes are processed.

\end{itemize}

The workflow of CORF-GS is shown in Fig.~\ref{fig:workflow}. 
Overall, CORF-GS uses fine-grained optical observations to efficiently guide fast Gaussian expansion and geometry initialization, allowing the maintained ORF to provide strong structural support for real-time incremental WRF reconstruction. Meanwhile, coupled optical-RF optimization further mitigates the discrepancy between optical and radio propagation and jointly improves the rendering quality of both modalities.

\subsection{Coupled Optical-RF Gaussian Representation}
Light and radio waves are both electromagnetic waves governed by Maxwell's equations. Although they have different wavelengths and material-dependent propagation behaviors, optical images and RF spectra are both generated by electromagnetic interactions with the same 3D environment, inherently constrained by common scene structures. 

Based on this observation, CORF-GS represents the ORF and WRF with coupled Gaussians, where the two modalities share Gaussian geometry while maintaining modality-specific attributes. Each primitive is parameterized as:

\begin{equation}
    \mathcal{G}_i^{\mathrm{c}} 
    =
    \{
    \boldsymbol{\mu}_i,
    \boldsymbol{\Sigma}_i,
    o_i^{\mathrm{opt}},
    c_i^{\mathrm{opt}},
    o_i^{\mathrm{rf}},
    c_i^{\mathrm{rf}}
    \},
\end{equation}
where $\boldsymbol{\mu}_i$ and $\boldsymbol{\Sigma}_i$ denote the shared Gaussian geometry, $\{o_i^{\mathrm{opt}}, c_i^{\mathrm{opt}}\}$ describe the optical appearance, while $\{o_i^{\mathrm{rf}}, c_i^{\mathrm{rf}}\}$ characterize the RF-specific response.

\subsection{Real-Time WRF Reconstruction}
\subsubsection{Optical-Guided Gaussian Expansion}
Upon the arrival of a new keyframe $\mathcal{K}_t$, CORF-GS first determines where new Gaussian primitives should be added. Instead of relying on gradient-based densification, we adopt a proactive sampling strategy~\cite{meuleman2025fly} focusing on informative and under-reconstructed areas. Since optical images provide reliable structural details, the primitive expansion is guided by optical images. RF spectra, as low-resolution angular power observations affected by multi-path superposition, are used for coupled optimization rather than direct sampling, avoiding ambiguous or redundant primitive insertion.

First, an initial probability map from the optical image $\mathbf{I}_t$ using the Laplacian-of-Gaussian (LoG) is computed as:
\begin{equation}
    P_t(\mathbf{u}) = \min \left(\left\| \nabla^2 \left( G_{\sigma} * \mathbf{I}_t
    \right)(\mathbf{u}) \right\|, 1
    \right),
\end{equation}
where $\mathbf{u}$ denotes a pixel coordinate and $G_{\sigma}$ is a Gaussian kernel with a standard deviation $\sigma$. This probability assigns higher sampling scores to high-frequency regions in the images, which usually indicate important scene structures.

Then, to avoid repeatedly inserting Gaussians into well-represented regions, a rendered optical image $\hat{\mathbf{I}}_t$ from the same pose is used to compute a penalty map:
\begin{equation}
    \tilde{P}_t(\mathbf{u})  = \min \left(\left\|\nabla^2  \left(
    G_{\sigma} * \hat{\mathbf{I}}_t \right)(\mathbf{u}) \right\|,1 \right).
\end{equation}
Consequently, the final sampling probability is defined as
\begin{equation}
    \bar{P}_t(\mathbf{u}) = \max \left( P_t(\mathbf{u}) - \tilde{P}_t(\mathbf{u}), 0 \right).
\end{equation}
Pixels are sampled according to $\bar{P}_t(\mathbf{u})$ to initialize new primitives, enabling efficient Gaussian expansion with structural awareness for real-time WRF reconstruction.

\subsubsection{Parameter Initialization}
After obtaining the sampled pixels, CORF-GS initializes the shared geometry and modality-specific attributes of each new primitive. For each sampled pixel $\mathbf{u}_i$, we first estimate its depth $z_i$~\cite{meuleman2025fly}. A monocular depth prior is obtained using Depth-Anything-2~\cite{yang2024depth} and is aligned to triangulated depths. The aligned monocular depth is then used as a prior for guided multi-view matching over neighboring keyframes, producing a refined depth estimate for sampled pixels. 
Given the refined depth $z_i$, the Gaussian center is initialized by back-projecting the sampled pixel into 3D space:
\begin{equation}
    \boldsymbol{\mu}_i = \Pi^{-1} \left( \mathbf{u}_i,
    z_i, \mathbf{P}_t  \right),
\end{equation}
where $\Pi^{-1}(\cdot)$ denotes the back-projection operation.

The scale of the new Gaussian is initialized from the sampling probability. Following the probability-based scale estimation~\cite{meuleman2025fly}, we first compute the expected nearest-neighbor distance in the image space:
\begin{equation}
    s_i^{\prime} = \frac{1} {2\sqrt{P_t(\mathbf{u}_i)}}.
\end{equation}
Then, convert $s_i^{\prime}$  into the 3D space by
\begin{equation}
    s_i = \frac{z_i s_i^{\prime}}{f},
\end{equation}
where $f$ denotes the focal length. Finally, the Gaussian is initialized with isotropic scaling using $s_i$.

For attributes, the optical and RF appearance are initialized from the RGB value of the sampled pixel and the synchronized RF spectrum, respectively. The opacity $o_i^{\mathrm{opt}}$ and $o_i^{\mathrm{rf}}$ are initialized according to the reliability of the newly sampled point, and are further optimized during subsequent training. 

\subsection{Optimization Pipeline}
\subsubsection{Loss Function}
After expanding Gaussian primitives for an incoming keyframe, CORF-GS selects a training keyframe $\mathcal{K}_{\tau} = \{\mathbf{I}_{\tau}, \mathbf{S}_{\tau}, \mathbf{P}_{\tau}\}$ from the active keyframe buffer for optimization, where $1 \le \tau \le t$. In our implementation, the latest keyframe is selected with probability 0.2, and otherwise one keyframe is sampled from the buffer. Given $\mathcal{K}_{\tau}$, CORF-GS renders the optical image $\hat{\mathbf{I}}_{\tau}$ and RF spectrum $\hat{\mathbf{S}}_{\tau}$ from the current ORF and WRF.

The optical reconstruction loss is defined as a combination of a $\mathcal{L}_1$ loss, a structural similarity index measure (SSIM) loss $1-\mathcal{L}_\mathrm{s}$, and a depth regularization term $\mathcal{L}_\mathrm{d}$:
\begin{equation}
    \mathcal{L}_{\mathrm{opt}} = (1-\gamma) \mathcal{L}_1\left(\hat{\mathbf{I}}_\tau,\mathbf{I}_\tau \right) + 
    \gamma \left( 1- \mathcal{L}_\mathrm{s} \left(\hat{\mathbf{I}}_\tau,\mathbf{I}_\tau \right) \right)+ \eta \mathcal{L}_\mathrm{d},
\end{equation}
where $\gamma$ and $\eta$ are weighting factors. The depth loss $\mathcal{L}_\mathrm{d}$ is computed between the rendered inverse depth and the aligned monocular inverse depth prior. Similarly, the RF reconstruction loss is computed between the rendered RF spectrum and the observed RF spectrum:
\begin{equation}
    \mathcal{L}_{\mathrm{rf}} = (1-\gamma) \mathcal{L}_1 \left(\hat{\mathbf{S}}_\tau,\mathbf{S}_\tau \right)  + \gamma \left( 1- \mathcal{L}_\mathrm{s}\left(\hat{\mathbf{S}}_\tau,\mathbf{S}_\tau \right) \right).
\end{equation}

\subsubsection{Coupled Optical-RF Optimization}
CORF-GS jointly optimizes the shared geometry $\{\boldsymbol{\mu},\boldsymbol{\Sigma}\}$ and the modality-specific appearance attributes $\{o^{\mathrm{opt}},c^{\mathrm{opt}}\}$ and $\{o^{\mathrm{rf}},c^{\mathrm{rf}}\}$ in the same optimization step. Let $\lambda_{\mathrm{rf}}$ denote the weighting factor that balances the ORF and WRF reconstruction objectives. Consequently, the overall training objective is:
\begin{equation}
    \mathcal{L} =  \mathcal{L}_{\mathrm{opt}} + \lambda_{\mathrm{rf}}  \mathcal{L}_{\mathrm{rf}},
\end{equation}

\begin{figure}[t]
    \centering
    \includegraphics[width=0.98\columnwidth]{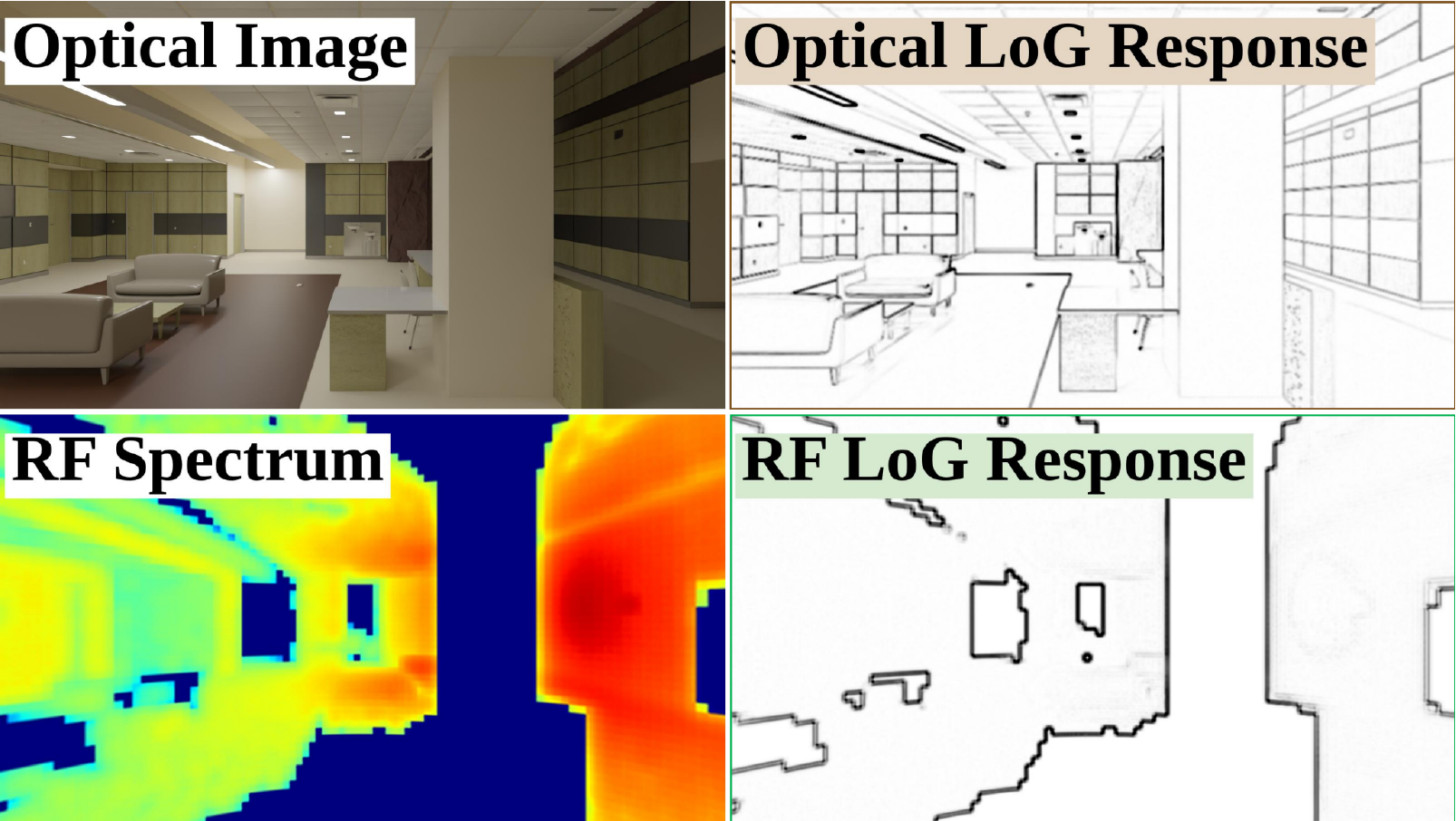}
    \caption{Optical-RF LoG response comparison. Optical LoG responses highlight fine-grained visual structures, whereas RF LoG responses reveal RF-informative regions that are not always aligned with optical edges, motivating coupled optical-RF optimization.}
    \label{fig:log_prob}
\end{figure}

This coupled optimization is crucial since the incrementally spawned Gaussians are initially guided by optical observations. Although images provide rich structural cues, the optically initialized Gaussian geometry may overlook RF-informative regions due to wavelength-dependent propagation discrepancies, as shown in Fig.~\ref{fig:log_prob}. Compared to two-stage optimization pipelines~\cite{zhang2026rf,wen2026bridging}, CORF-GS allows RF spectra to adjust the shared geometry by coupling optical and RF supervision, preventing the WRF from passively adapting to a frozen optical geometry.

\subsubsection{Training Strategy}
To improve training efficiency, CORF-GS adopts a coarse-to-fine training strategy~\cite{huang2024photo, meuleman2025fly}. Both optical images and RF spectra are downsampled by a factor of $2^l$ to form multi-scale pyramids, and are optimized from coarse to fine levels. This enables efficient global structure learning followed by detailed texture refinement. Additionally, after all $T$ keyframes are processed, an optional offline global fine-tuning stage can be applied to further improve reconstruction accuracy.

\section{Experiments}
\subsection{Experimental Setups}
\subsubsection{Dataset}
We evaluate CORF-GS on an open-source optical-RF dataset~\cite{zhang2026rf}. This dataset is built upon a digital replica of a $14 \times 15$~m$^2$ lobby at the National Institute of Standards and Technology (NIST). The optical images are rendered using Blender, while the RF spectra are simulated via Sionna RT~\cite{hoydis2023sionna}. The carrier frequency is set to 60~GHz. The TX is equipped with a single antenna, while the RX utilizes an $8\times8$ planar antenna array. To construct spatial spectra, multi-path components are first mapped to angular domain and subsequently reorganized using a Jet color map and aligned with a pinhole camera projection.

To support the evaluation of real-time WRF reconstruction, we structure the data acquisition into a sequential format. During dataset collection, the TX remains stationary while the camera-RX platform moves continuously along a predefined trajectory that loops around the lobby. This process yields a sequence of 147 keyframes in total. We reserve one keyframe every eight frames for evaluation, resulting in 19 test keyframes, and use the remaining 128 keyframes for training. For more details, please refer to Appendix~A2.

\begin{figure*}[t]
    \centering
    \includegraphics[width=0.98\textwidth]{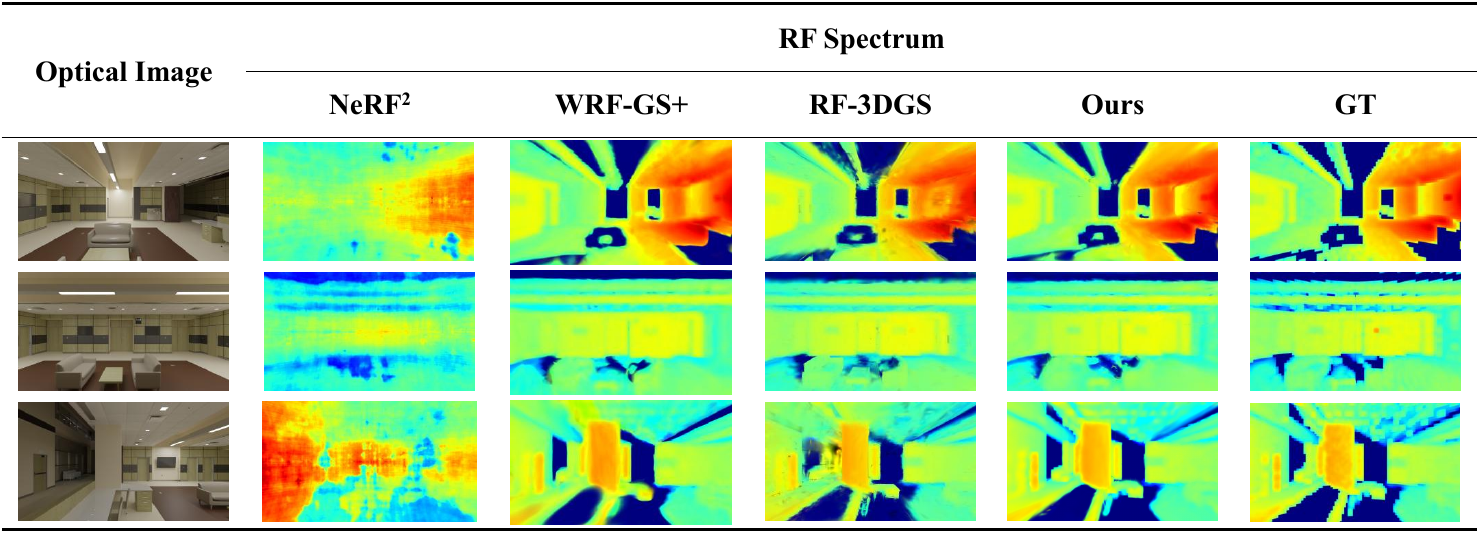}
    \caption{Qualitative results of novel spectrum synthesis after WRF reconstruction. CORF-GS better matches the ground-truth spectra in both global energy distribution and local high-frequency structures compared with baselines.}
    \label{fig:baseline-results}
\end{figure*}

\begin{table}[t]
  \centering

  \setlength{\tabcolsep}{10pt}
  \renewcommand{\arraystretch}{1.2}
  \caption{Quantitative comparison of WRF reconstruction efficiency and novel spectrum synthesis quality. The total time of RF-3DGS is the sum of the runtimes of its two stages. CORF-GS achieves the best rendering performance while requiring less reconstruction time than the offline baselines.}
  \begin{tabular}{l|cccc}
    \toprule
    Method & PSNR${_\mathrm{rf}}^\uparrow$ & SSIM${_\mathrm{rf}}^\uparrow$ & LPIPS${_\mathrm{rf}}^\downarrow$ & Time$^\downarrow$\\
    \midrule
    NeRF$^2$ & 12.17 & 0.594 & 0.527 & 4:09:01 \\
    WRF-GS+  & 17.44 & 0.839 & 0.252 & 0:42:20 \\
    RF-3DGS & 16.92 & 0.788 & 0.333 & 0:14:41 \\
    Ours & \textbf{18.52} & \textbf{0.852} & \textbf{0.239} & \textbf{0:02:18} \\
    \bottomrule
  \end{tabular}
  \label{tab:baseline-results-metric}
\end{table}

\subsubsection{Metrics and Baselines}
We evaluate the synthesis quality using peak signal-to-noise ratio (PSNR), SSIM, and learned perceptual image patch similarity (LPIPS)~\cite{zhang2018unreasonable}. Metrics for optical and RF are designated with subscripts $\mathrm{opt}$ and $\mathrm{rf}$, respectively. Besides, optimization time is reported to evaluate the real-time performance.

The WRF reconstruction baselines include the following:
\begin{itemize}
    \item \textbf{NeRF$^2$}~\cite{zhao2023nerf2}: NeRF\(^2\) reconstructs the WRF using an implicit neural representation and predicts RF spectra through volume rendering. It represents a NeRF-based approach for WRF reconstruction.

    \item \textbf{WRF-GS+}~\cite{wen2025neural}: WRF-GS+ is a 3DGS-based WRF reconstruction method. It leverages deformable 3D Gaussians to model both static and dynamic components of the WRF.

    \item \textbf{RF-3DGS}~\cite{zhang2026rf}: RF-3DGS follows a two-stage 3DGS-based spectrum synthesis pipeline. It first reconstructs the optical Gaussian geometry and then freezes the optical-stage geometry to optimize RF-related attributes for WRF reconstruction. 
\end{itemize}

To ensure fair comparison, we adapt WRF-GS+ and NeRF$^2$ to our spatial spectrum dataset. CORF-GS processes optical-RF keyframes sequentially, whereas all baselines reconstruct WRF offline using the full set of observations assumed to be available beforehand.

\subsubsection{Implementation Details}
CORF-GS is trained and tested on an NVIDIA GeForce RTX 4090 GPU with an Intel Core i9-14900KF CPU. We set the SSIM loss weight $\gamma$ to 0.2, the initial depth regularization weight $\eta$ to 0.01 with a decay factor of 0.9 at each iteration, and the RF loss weight $\lambda_{\mathrm{rf}}$ to 0.1 (see Appendix~A1 for the sensitivity analysis of $\lambda_{\mathrm{rf}}$). The number of pyramid levels is set to 3. 
For each keyframe, CORF-GS performs 75 optimization iterations. Each experiment is repeated ten times, and we report the averaged result. For the offline baselines, WRF-GS+ and NeRF$^2$ are optimized for 10,000 and 30,000 iterations, respectively. For RF-3DGS, the optical and RF stages are optimized for 30,000 and 10,000 iterations, respectively. All other configurations of the baseline methods follow their original settings.

\subsection{Main Results}
\begin{table}[!t]
  \centering
  \renewcommand{\arraystretch}{1.2}
  \caption{Comparison of novel spectrum synthesis quality. \\ For RF-3DGS, $(N_{\mathrm{opt}},N_{\mathrm{rf}})_{\times 10^3}$ denotes the optical-stage and RF-stage training iterations, respectively.}
  \begin{tabular}{lc|ccc}
    \toprule
    Method & Iterations & PSNR${_\mathrm{rf}}^\uparrow$ & SSIM${_\mathrm{rf}}^\uparrow$ & LPIPS${_\mathrm{rf}}^\downarrow$ \\
    \midrule
    \multirow{4}{*}{WRF-GS+} & 5$_{\times 10^3}$ & 16.75 & 0.826 & 0.271\\
    & 10$_{\times 10^3}$ & 17.44 & 0.839 & 0.252\\
    & 20$_{\times 10^3}$ & 17.96 & 0.848 & 0.243 \\
    & 30$_{\times 10^3}$ & 18.09 & 0.850 & 0.243 \\

    \midrule
    & (15,10)$_{\times 10^3}$ & 17.19 & 0.792 & 0.320 \\
    &  (15,25)$_{\times 10^3}$ & 17.70 &0.807 & 0.307 \\
    RF-3DGS &  (15,40)$_{\times 10^3}$ & 17.88 &0.812 & 0.303 \\
    & (30,10)$_{\times 10^3}$ & 16.92 & 0.788 & 0.333 \\
    & (30,25)$_{\times 10^3}$ & 17.38 & 0.804 & 0.319 \\

    \midrule
    \multirow{4}{*}{Ours} & 50 & 18.18 &	0.843 &	0.247\\
     & 75 & 18.52 & 0.852 & 0.239\\
     & 100 & 18.56 & \textbf{0.854} & 0.237\\
     & 125 & \textbf{18.65} & \textbf{0.854} & \textbf{0.236}\\
    \bottomrule
  \end{tabular}
  \label{tab:baseline-iter}
\end{table}

\begin{figure}[t]
    \centering
    \includegraphics[width=0.98\columnwidth]{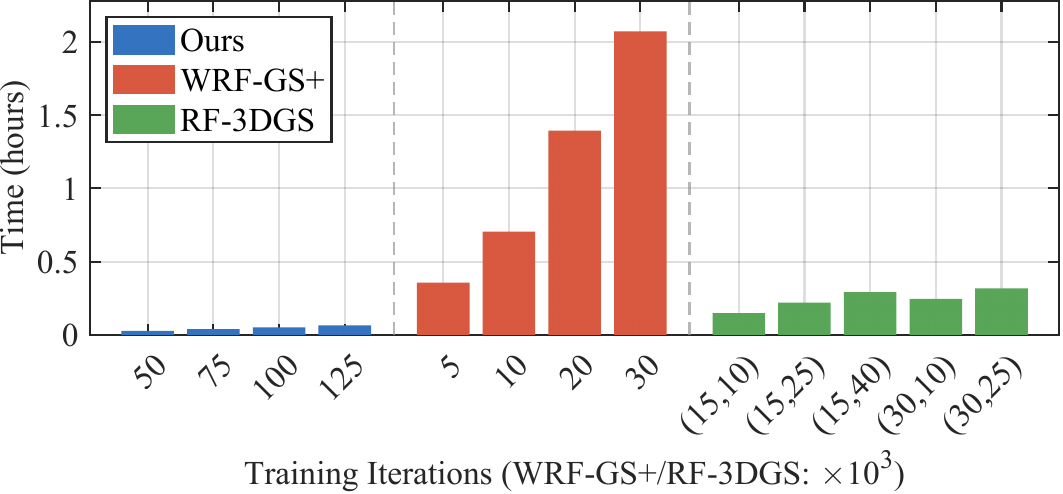}
    \caption{Comparison of time costs with various iterations.}
    \label{fig:baseline-iter-time}
\end{figure}

\begin{table*}[t]
  \centering
  \renewcommand{\arraystretch}{1.25}
  \caption{Novel view synthesis quality and total optimization time of CORF-GS with additional offline fine-tuning iterations.}
  \begin{tabular}{l|cccccccc}
    \toprule
    Iterations & PSNR${_\mathrm{opt}}^\uparrow$ & SSIM${_\mathrm{opt}}^\uparrow$ & LPIPS${_\mathrm{opt}}^\downarrow$ & PSNR${_\mathrm{rf}}^\uparrow$ & SSIM${_\mathrm{rf}}^\uparrow$ & LPIPS${_\mathrm{rf}}^\downarrow$ & Time (s)$^\downarrow$\\
    \midrule
    25 & 38.13 & 0.980 & 0.069 & 18.85 & 0.862 & 0.229 & \textbf{191.69} \\
    50 & 38.30 & 0.980	& 0.066 & 18.96 & 0.865 & 0.226 & 243.29 \\
    75 & \textbf{38.63} & \textbf{0.982} & 0.064 & 19.04 & 0.867 & \textbf{0.223} & 295.47 \\
    100 & 38.60 & \textbf{0.982} & 0.063 & \textbf{19.09} &	\textbf{0.868} & \textbf{0.223} &	363.14 \\
    \midrule
    RF-3DGS & 38.39 & 0.980 & \textbf{0.062} & 16.92 & 0.788 & 0.333 & 881.00 \\
    \bottomrule
  \end{tabular}

  \label{tab:finetune}
\end{table*}

\begin{table*}[t]
  \centering
  \renewcommand{\arraystretch}{1.25}
  \caption{Ablation studies. The variant w/o optical-guided exp. replaces the optical-guided Gaussian expansion with uniform sampling. The variant w/o joint (frozen geo.) adopts two-stage training and freezes the optically optimized Gaussian geometry during RF optimization, while w/o joint (RF-tuned geo.) allows further refining the geometry in the second stage.}
  \begin{tabular}{l|ccccccc}
    \toprule
    Ablation & PSNR${_\mathrm{opt}}^\uparrow$ & SSIM${_\mathrm{opt}}^\uparrow$ & LPIPS${_\mathrm{opt}}^\downarrow$ & PSNR${_\mathrm{rf}}^\uparrow$ & SSIM${_\mathrm{rf}}^\uparrow$ & LPIPS${_\mathrm{rf}}^\downarrow$ & Time$^\downarrow$\\
    \midrule
    w/o optical-guided exp. & 31.89 & 0.966 & 0.112 & 15.69 & 0.780 & 0.317 & 0:07:15 \\
    w/o joint (frozen geo.) & \textbf{37.23} & \textbf{0.979} & \textbf{0.076} & 17.11 & 0.798 & 0.297 & \textbf{0:02:13} \\
    w/o joint (RF-tuned geo.) & 36.72 & 0.977 & 0.080 & 18.44 & 0.836 & 0.240 & 0:02:31 \\
    OursFull & 37.14 & 0.978 & 0.077 & \textbf{18.52} & \textbf{0.852} & \textbf{0.239} & 0:02:18 \\
    \bottomrule
  \end{tabular}
  
  \label{tab:ablation}
\end{table*}

\subsubsection{Comparison with Baselines}
As shown in Fig.~\ref{fig:baseline-results}, CORF-GS produces RF spectra that are more consistent with the ground truth. In contrast, NeRF$^2$ struggles to recover the overall power distribution. Quantitatively, Table~\ref{tab:baseline-results-metric} shows that CORF-GS achieves the best performance, with PSNR$_\mathrm{rf}$ of 18.52~dB, SSIM$_\mathrm{rf}$ of 0.852, and LPIPS$_\mathrm{rf}$ of 0.239. CORF-GS requires only about 2 minutes for WRF reconstruction, considerably faster than the baselines. These results show that CORF-GS improves spectrum synthesis quality while significantly reducing the reconstruction time.

\subsubsection{Performance with Various Training Costs}
Table~\ref{tab:baseline-iter} and Fig.~\ref{fig:baseline-iter-time} analyze the trade-off between training cost and performance. CORF-GS achieves competitive spectrum synthesis performance with only a few iterations.
However, even with $30\times 10^3$ iterations, WRF-GS+ only reaches PSNR$_\mathrm{rf}$ of 18.09~dB and costs more than two hours for WRF reconstruction. 
For RF-3DGS, more RF-stage iterations improve spectrum synthesis, but the performance remains below CORF-GS under all tested configurations. 
Moreover, comparing $(15,10)_{\times10^3}$ with $(30,10)_{\times10^3}$ shows that more optical-stage iterations do not necessarily improve WRF reconstruction, suggesting that optically optimized geometry may be suboptimal for RF modeling.

\subsubsection{Performance with Extra Fine-tuning}
Table~\ref{tab:finetune} reports the performance of CORF-GS under different numbers of fine-tuning iterations. As the number of fine-tuning iterations increases, the quality of both optical image and RF spectrum synthesis improves, while the RF performance consistently remains better than that of RF-3DGS. With only 75 additional fine-tuning iterations, CORF-GS achieves comparable optical reconstruction quality to RF-3DGS, while requiring only about 33\% of its reconstruction time. These results indicate that fine-tuning provides an optional accuracy-efficiency trade-off. It can be applied when higher ORF fidelity is required, while the real-time stage already achieves superior WRF performance with low reconstruction time.

\subsection{Ablation Study}
\subsubsection{Effect of Optical-Guided Gaussian Expansion}
We replace the structure-aware sampling with uniform sampling, where each image pixel has a probability of 0.1 to spawn a new Gaussian. As summarized in Table~\ref{tab:ablation}, uniform sampling substantially degrades both optical and RF rendering. We also observe that uniform sampling produces more Gaussians, increasing the final primitive count from about 0.2M to 1.0M, accompanied by about $3.1\times$ longer reconstruction time. These results indicate that the optical-guided sampling effectively captures structural details, which supports more efficient expansion and better reconstruction quality.

\subsubsection{Effect of Coupled Optical-RF Optimization}
In the variant \textit{w/o joint (frozen geo.)}, the coupled optical-RF optimization is replaced by a two-stage pipeline as in RF-3DGS. Experimental results show that while this two-stage strategy preserves ORF reconstruction quality, the spectrum synthesis performance degrades significantly, with PSNR$_\mathrm{rf}$ dropping from 18.52~dB to 17.11~dB. Therefore, an exclusively optically optimized Gaussian geometry is insufficient to support WRF reconstruction. When the scene structure is frozen, the RF spectra only passively fit the pre-existing Gaussian geometry, limiting the representation of RF-informative regions.

\subsubsection{Effect of Two-Stage Gaussian Geometry Refinement}
The previous experiment shows that a completely frozen, optically optimized Gaussian geometry restricts WRF reconstruction. Compared with \textit{w/o joint (frozen geo.)}, \textit{w/o joint (RF-tuned geo.)} further examines whether allowing RF supervision to fine-tune the Gaussian geometry in the second stage can replace coupled optical-RF optimization. As shown in Table~\ref{tab:ablation}, RF-stage geometry refinement improves PSNR$_\mathrm{rf}$ from 17.11~dB to 18.44~dB, confirming the effectiveness of RF supervision for geometry refinement. While geometry refinement is beneficial, it still underperforms the full model, indicating that coupled optimization remains more effective.

\section{Conclusion}
In this paper, we propose CORF-GS, a 3DGS-based real-time WRF reconstruction framework from sequential optical-RF keyframes.
In this framework, we leverage the fine-grained structural details in images to assist WRF reconstruction.
CORF-GS builds a unified Gaussian representation, where optical and RF modalities share the same Gaussian geometry while maintaining modality-specific appearance attributes. 
During the real-time reconstruction process, CORF-GS first employs an optical-guided Gaussian expansion strategy to sample new primitives in under-represented regions. Subsequently, to mitigate the physical response mismatch between light and radio waves, a coupled optical-RF optimization is introduced to jointly refine the shared Gaussian representation. Experiments show that CORF-GS achieves state-of-the-art RF spectrum synthesis quality with PSNR$_\mathrm{rf}=18.52$ dB, while reducing the reconstruction time to only $2$ min $18$ s, significantly faster than existing WRF methods.

\bibliographystyle{IEEEtran}
\bibliography{IEEEabrv,reference}

\end{document}